\documentclass[11pt,a4paper]{article}

\usepackage[margin=1in]{geometry}
\usepackage{amsmath,amssymb}
\usepackage{graphicx}
\usepackage[font=small]{caption}
\usepackage{authblk}
\usepackage{xcolor}
\usepackage[hidelinks]{hyperref}

\title{Josephson energy of superconducting junctions: amorphous versus crystalline tunnel barriers}

\author[1]{Wanting Zhang}
\author[1]{Aldilene Saraiva-Souza}
\author[1]{F\'elix Beaudoin}
\author[2]{Xianghua Kong}
\author[3]{Hong Guo}
\author[1]{Yu Zhu\thanks{Corresponding author: \href{mailto:eric@nanoacademic.com}{eric@nanoacademic.com}}}

\affil[1]{Nanoacademic Technologies Inc., Montr\'eal, QC H3A 1E7, Canada}
\affil[2]{College of Physics and Optoelectronic Engineering, Shenzhen University, Shenzhen 518060, China}
\affil[3]{Department of Physics, McGill University, Montr\'eal, QC H3A 0G4, Canada}

\date{\today}

\begin{document}

\maketitle

\begin{abstract}
The Josephson energy $E_J$ is a key parameter governing the properties of transmon superconducting qubits. In Al/AlO$_x$/Al junctions, $E_J$ is set by electron tunneling through an ultrathin oxide barrier and therefore depends exponentially on the atomic structure of the barrier. We compute $E_J$ by first-principles device modeling based on the NEGF-DFT quantum-transport method, comparing a junction with a crystalline Al$_2$O$_3$ barrier against ten junctions with melt--quenched amorphous Al$_2$O$_3$ barriers of the same thickness. From the Fermi-level transmission and the Ambegaokar--Baratoff relation, we obtain a mean $E_J/h$ of $2.78$~GHz for the amorphous ensemble, with a standard deviation of $4.67$~GHz, compared with $0.73$~GHz for the crystalline reference; individual amorphous values span nearly two orders of magnitude. Scattering-state analysis shows that transport is quantum tunneling and that the variability originates from stoichiometric inhomogeneity of the amorphous oxide: Al-rich, low-barrier regions can connect into percolation-like tunneling pathways that strongly enhance the conductance. A realistic $200\times200$~nm$^2$ junction self-averages over more than $2\times10^4$ such microscopic regions. These results establish a quantitative atomistic route from oxide microstructure to the superconducting-circuit energy scale $E_J$.
\end{abstract}


\section{Introduction}
\label{sec:introduction}

Josephson junctions provide the nonlinear inductance at the heart of superconducting quantum circuits. In the transmon qubit, the circuit is described by the Hamiltonian~\cite{koch2007,krantz2019}
\begin{equation}
	\hat{H} = 4E_C\left(\hat{n}-n_g\right)^2 - E_J\cos\hat{\varphi},
	\label{eq:transmon}
\end{equation}
where $\hat{n}$ is the Cooper-pair number operator, $n_g$ the offset charge, $\hat{\varphi}$ the phase difference across the junction, $E_C$ the charging energy, and $E_J$ the Josephson energy. Together, $E_J$ and $E_C$ set the qubit frequency and anharmonicity, making $E_J$ one of the key parameters governing superconducting quantum circuits.

Superconducting quantum circuits exhibit a striking separation of length scales: readout resonators are millimeters long and transmon capacitor pads measure hundreds of micrometers~\cite{krantz2019}, whereas the AlO$_x$ tunnel barrier is only 1--2~nm thick. Since the junction is far smaller than every other circuit element, it is common practice to model it by the two lumped parameters $E_J$ and $E_C$ taken from experimental calibration. On the other hand, $E_J$ arises from quantum tunneling and thus depends \emph{exponentially} on the barrier thickness and height, which are determined by atomic-scale details such as the Al/AlO$_x$ interface structure, hydrogen contamination, and structural defects. A full understanding of $E_J$ therefore requires an atomistic model of the junction.

Such a model is obtained by combining density functional theory (DFT) with the nonequilibrium Green's function formalism (NEGF)~\cite{taylor2001}: DFT captures the material properties arising from the atomic and electronic structure, while NEGF describes the quantum transport through an open two-probe system, in particular tunneling. The workflow to extract $E_J$ for an Al/AlO$_x$/Al junction---two Al electrodes separated by an ultrathin oxide barrier---proceeds in four steps. First, the energy-dependent transmission coefficient $T(E)$ is computed via NEGF-DFT; because of the tunneling process, its value at the Fermi energy, $T(E_F)$, depends exponentially on the atomic details of the barrier. Second, $T(E_F)$ gives the normal-state conductance of the simulation cell of transverse area $A_0$ via the Landauer formula, which is scaled to the device area $A$ and inverted to give the normal-state resistance. Because the crystalline and amorphous junctions employ different laterally periodic supercells, $A_0$ is evaluated separately for each junction geometry:
\begin{equation}
	G_N^{A_0} = G_0\,T(E_F), \quad
	G_N^{A} = G_N^{A_0}\,\frac{A}{A_0}, \quad
	R_N^{A} = \frac{1}{G_N^{A}},
	\label{eq:landauer}
\end{equation}
with $G_0 = 2e^2/h$. Third, the critical current $I_c$ follows from the zero-temperature Ambegaokar--Baratoff relation~\cite{ambegaokar1963}
\begin{equation}
	I_c R_N^{A} = \frac{\pi\Delta}{2e},
	\label{eq:ab}
\end{equation}
where $\Delta$ is the superconducting energy gap. Finally,
\begin{equation}
	E_J = \frac{\hbar I_c}{2e}.
	\label{eq:ej}
\end{equation}
The chain $T(E_F) \rightarrow R_N \rightarrow I_c \rightarrow E_J$ connects the microscopic barrier structure directly to the Josephson energy $E_J$, a central energy scale of superconducting quantum circuits.

Experimental characterization shows that real AlO$_x$ barriers are structurally and chemically nonuniform: transmission electron microscopy reveals barrier-thickness variations and oxygen deficiency near the metal--oxide interfaces~\cite{zeng2016}, and the recent observation of higher Josephson harmonics indicates that practical tunnel junctions contain conduction channels with a distribution of transparencies rather than a single uniform one~\cite{willsch2024}. These findings motivate atomistic modeling that connects the atomic arrangement of the oxide to the electronic transmission through the junction.

A substantial body of atomistic work has developed along these lines. Koberidze \textit{et al.} used DFT to determine how Al/Al$_2$O$_3$ interface geometry, stacking, and termination shape the tunnel-barrier profile~\cite{koberidze2016,koberidze2018}. DuBois \textit{et al.} constructed ab initio models of ultrathin Al--AlO$_x$--Al barriers~\cite{dubois2016}, and Cyster \textit{et al.} simulated the oxidation-based fabrication of aluminum-oxide tunnel junctions with molecular dynamics~\cite{cyster2021} and computed NEGF transport through three-dimensional atomistic AlO$_x$ junctions, finding that local variations in oxide density and stoichiometry can produce localized conduction channels even at uniform thickness~\cite{cyster2020}. Lapham and Georgiev combined DFT and NEGF to study oxide stoichiometry and device variability in Al/AlO$_x$/Al junctions~\cite{lapham2022}. Within the NEGF-DFT framework employed in the present work, junctions with \emph{crystalline} tunnel barriers have been examined in detail: Shan \textit{et al.} showed that an O-terminated interface makes transport insensitive to barrier thickness~\cite{shan2022}, Qiu \textit{et al.} examined how blurred (intermixed) interfaces modify junction transport~\cite{qiu2024}, and Fan \textit{et al.} studied the influence of the alumina stoichiometric ratio~\cite{fan2025}. These works demonstrate the strong sensitivity of transmission to interface termination, intermixing, and composition.

The main challenge in modeling realistic junctions is that the AlO$_x$ barrier is amorphous rather than crystalline: an amorphous barrier requires a much larger simulation cell and introduces stochastic variations in the atomic configuration, so a single structure is no longer representative. In this work we construct junctions with both crystalline and amorphous Al$_2$O$_3$ barriers of the same oxide thickness and compare their transport properties within the same first-principles device-modeling framework, extracting the Josephson energy for each sample. We find a mean $E_J/h$ of $2.78$~GHz for the ten amorphous junctions, with a standard deviation of $4.67$~GHz. The individual values span nearly two orders of magnitude around the crystalline reference of $0.73$~GHz, and we trace this variability to percolation-like tunneling pathways created by stoichiometric inhomogeneity in the amorphous oxide.

\section{Method}
\label{sec:method}

\subsection{Atomic structures}
\label{subsec:structures}

\begin{figure}[!htb]
	\centering
	\includegraphics[width=0.92\linewidth]{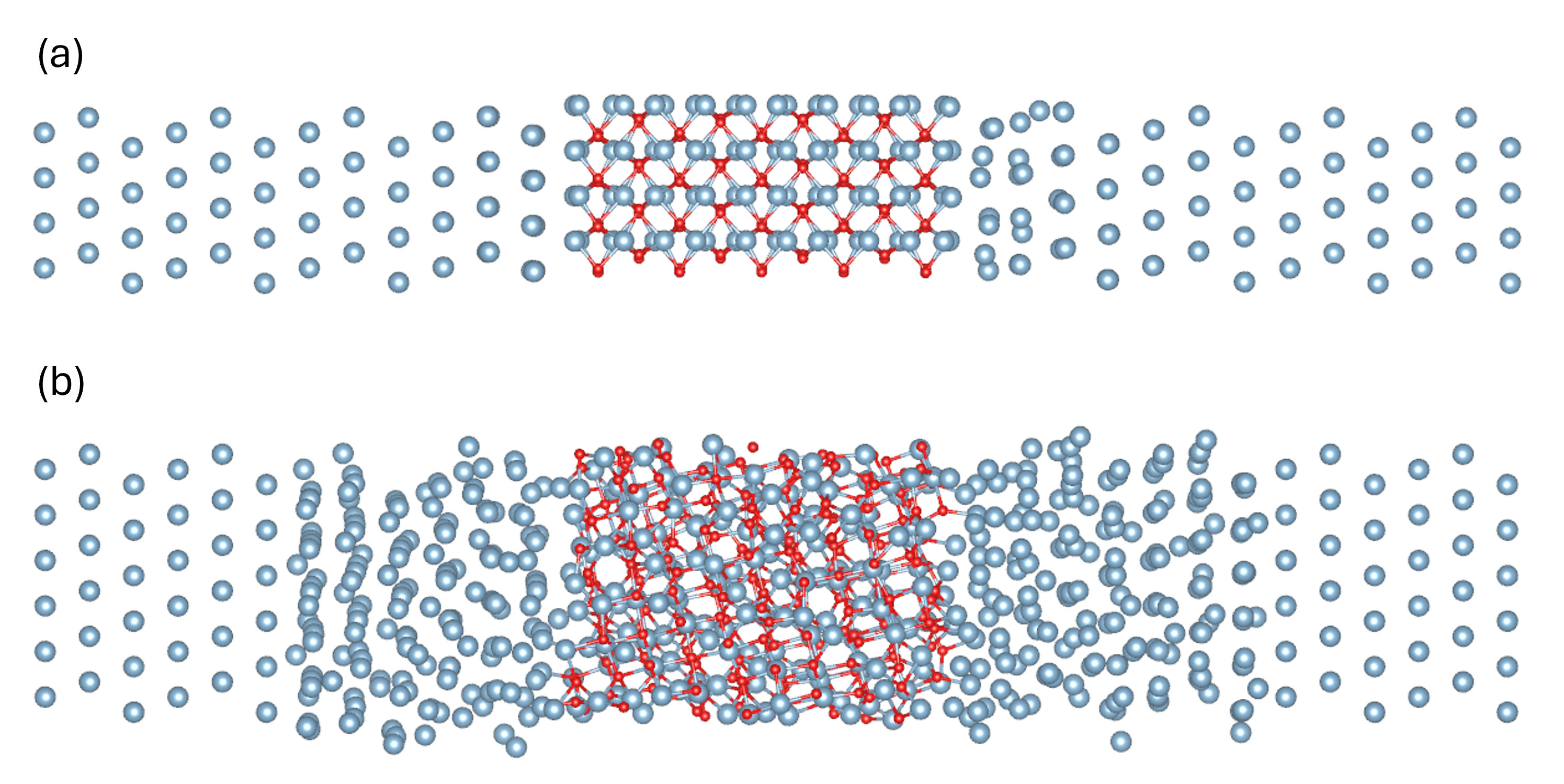}
	\caption{
		Atomic structures of the Josephson-junction models.
		(a) Junction with a crystalline Al$_2$O$_3$ tunnel barrier.
		(b) Junction with an amorphous Al$_2$O$_3$ tunnel barrier of the same oxide thickness.
		Blue and red spheres represent Al and O atoms, respectively.
	}
	\label{fig:atoms}
\end{figure}

Figure~\ref{fig:atoms} shows the two junction models studied here: a crystalline Al(111)--Al$_2$O$_3$--Al(111) junction and a representative amorphous junction with the same oxide thickness. The crystalline junction contains 420 atoms; each amorphous junction contains 1080 atoms.

\paragraph{Crystalline junction.}
The crystalline junction was built by connecting Al(111) slabs to an $\alpha$-Al$_2$O$_3$ barrier using a laterally commensurate supercell, so the Al and oxide regions share identical in-plane periodicity within the constructed junction. The transverse dimensions were chosen to preserve this periodic commensurability and therefore differ from those of the amorphous junctions. Enforcing the same transverse dimensions for the crystalline and amorphous models would not preserve the commensurate periodic structure and would require additional artificial strain or structural distortion. We reshaped the Al lattice to match the Al$_2$O$_3$ cell, rather than the reverse, to keep the electronic structure of the oxide unchanged: tunneling is much less sensitive to the metal band structure, for which mainly the density of states at the Fermi level matters. The oxide is terminated asymmetrically, with an Al-terminated surface on one side and an O-terminated surface on the other. The junction cell measured $9.61 \times 8.32 \times 80.13$~\AA$^3$, with an oxide-cell length of $19.67$~\AA. The structure was relaxed with the six outermost Al(111) layers on each side fixed, using the FIRE optimizer until the maximum force fell below $0.01$~eV~\AA$^{-1}$.

\paragraph{Amorphous junctions.}
The amorphous junctions were constructed in two steps. First, amorphous Al$_2$O$_3$ was generated from a crystalline $\alpha$-Al$_2$O$_3$ cell (405 atoms, $14.42 \times 12.48 \times 19.67$~\AA$^3$) by a melt--quench molecular-dynamics protocol in LAMMPS~\cite{thompson2022}. Interatomic interactions were described by the Matsui force field~\cite{matsui1994}, a long-range Coulomb term plus a short-range Buckingham potential optimized for bulk Al$_2$O$_3$, following its application to amorphous alumina by Pugliese \textit{et al.}~\cite{pugliese2022}. All runs used the NVT ensemble at fixed volume with a $0.5$~fs time step: the system was heated from $300$ to $5000$~K over $100$~ps, held at $5000$~K for $400$~ps, cooled to $4000$~K over $100$~ps and held for $200$~ps, quenched to $300$~K over $500$~ps, and equilibrated for $200$~ps ($1.5$~ns per trajectory). The procedure was repeated with ten random seeds, producing ten independent amorphous realizations.

Second, each amorphous oxide was inserted between two crystalline Al(111) slabs (12 layers on the left and 13 on the right, making the fcc stacking at the two interfaces equivalent) to form a two-probe structure, which was relaxed with the machine-learning interatomic potential CHGNet~\cite{deng2023}. CHGNet is trained on a large and diverse materials database, making it more reliable than single-material force fields for the heterogeneous metal--oxide interfaces present here. The six outermost Al layers on each side were fixed, and the BFGS optimizer was run until the maximum force fell below $0.02$~eV~\AA$^{-1}$. The relaxed junction has an in-plane cross section of $14.415 \times 12.484$~\AA$^2$ ($\approx 1.44 \times 1.25$~nm$^2$). Thus, the crystalline and amorphous transport cells have different transverse areas; this difference is explicitly accounted for in the transport analysis using the corresponding simulation-cell area $A_0$.

\subsection{Quantum transport calculations}
\label{subsec:transport}

Transport was computed with the NEGF-DFT first-principles method~\cite{taylor2001}, in which DFT incorporates the atomic details of the junction into the device Hamiltonian, while NEGF provides the quantum transport theory of the open system, describing scattering states and, in particular, tunneling. Each relaxed junction forms the central scattering region of a two-probe system connected to semi-infinite crystalline Al electrodes. The retarded Green's function of the central region is
\begin{equation}
	G^r(E) = \left[ES - H - \Sigma^r_L(E) - \Sigma^r_R(E)\right]^{-1},
	\label{eq:greens}
\end{equation}
where $H$ and $S$ are the Hamiltonian and overlap matrices and $\Sigma^r_{L,R}$ are the electrode self-energies, and the transmission coefficient is
\begin{equation}
	T(E) = \mathrm{Tr}\left[\Gamma_L(E)\,G^r(E)\,\Gamma_R(E)\,G^a(E)\right],
	\qquad
	\Gamma_{L,R} = i\left[\Sigma^r_{L,R}-\Sigma^a_{L,R}\right],
	\label{eq:transmission}
\end{equation}
with $G^a$ the advanced Green's function and $\Gamma_{L,R}$ the electrode couplings.

The calculations were performed with the NanoDCAL quantum-transport package~\cite{nanodcal,taylor2001}, which implements NEGF-DFT using a linear combination of atomic orbitals (LCAO) basis and norm-conserving pseudopotentials. NanoDCAL carries out the self-consistent field iteration in real space within the Keldysh NEGF framework, handles open two-probe boundary conditions, and provides transmission coefficients, scattering states, and their real-space projections; its parallelized implementation makes quantum transport feasible for junction models containing thousands of atoms. For both Al and O, the LCAO basis contained 13 functions per atom: two \(s\)-type radial functions, two \(p\)-type radial functions, and one \(d\)-type radial function; the compact LCAO basis reduces the computational cost dramatically compared with plane waves and enables transport calculations for systems with $10^3$ atoms. Exchange and correlation were treated at the GGA level, which tends to underestimate the oxide band gap, so absolute transmissions may be slightly overestimated; the trends across samples, however, should be robust.

\section{Results}
\label{sec:results}

\subsection{Josephson energies}
\label{subsec:ej}

\begin{figure}[!htb]
	\centering
	\includegraphics[width=0.7\linewidth]{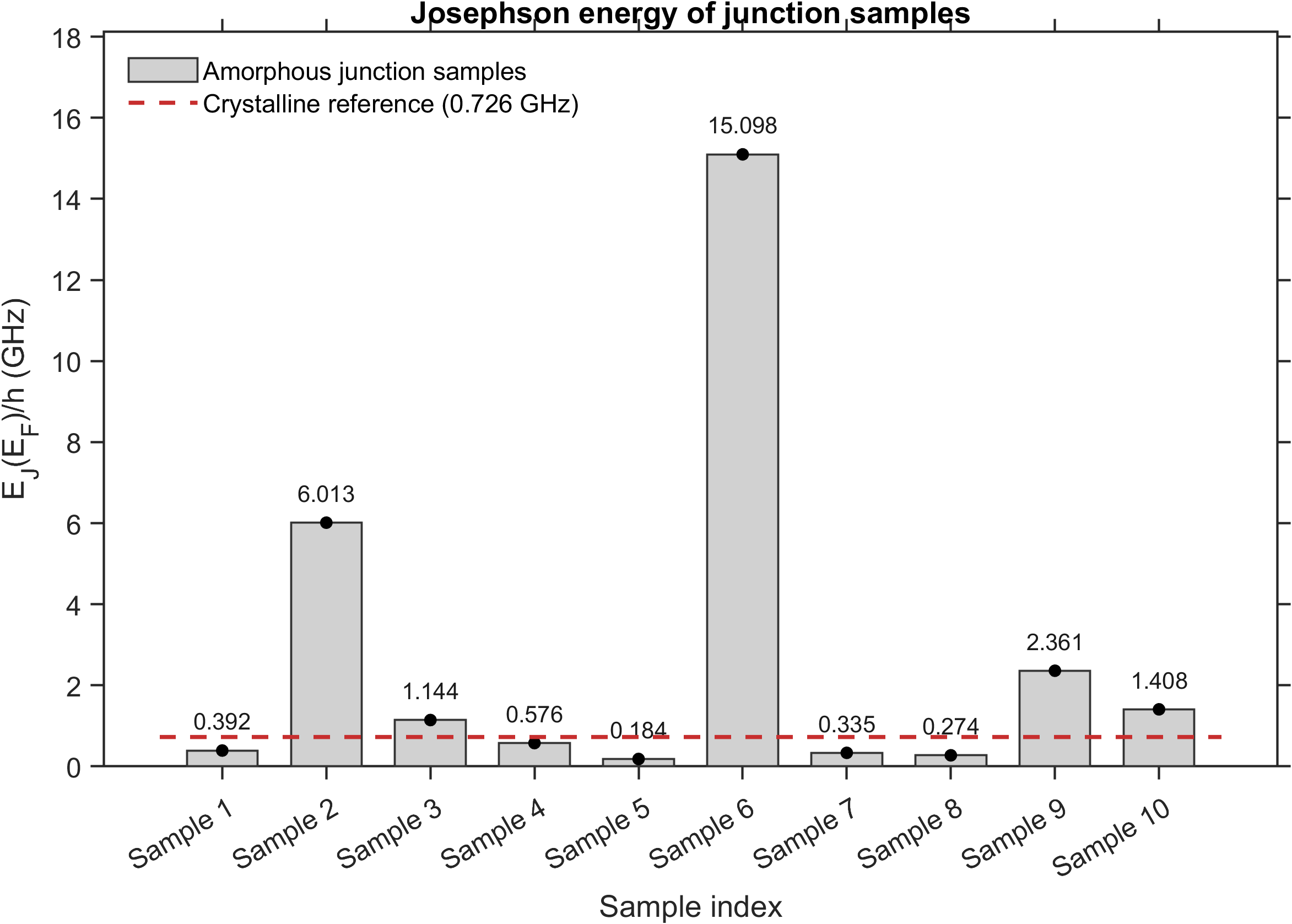}
	\caption{
		Josephson energies of the ten amorphous junction samples (bars), estimated from the calculated Fermi-level transmission using the Ambegaokar--Baratoff relation with $\Delta = 0.18$~meV and a device area of $200\times200$~nm$^2$. Each junction is scaled using its corresponding transverse simulation-cell area $A_0$.
		The dashed line marks the crystalline reference junction ($E_J/h \approx 0.73$~GHz).
	}
	\label{fig:ej}
\end{figure}

Figure~\ref{fig:ej} shows the Josephson energies of the ten amorphous junctions obtained from Eqs.~(\ref{eq:landauer})--(\ref{eq:ej}), using a superconducting gap $\Delta = 0.18$~meV and scaling each atomistic junction to a representative device area $A = 200\times200$~nm$^2$. The scaling uses the transverse simulation-cell area $A_0$ of each junction separately, thereby accounting for the different lateral dimensions of the crystalline and amorphous models. The dashed line marks the crystalline reference, for which $T(E_F) = 1.334\times10^{-6}$, giving $E_J/h \approx 0.73$~GHz.

For the ten amorphous junctions, we obtain a mean $E_J/h$ of $2.78$~GHz, with a standard deviation of $4.67$~GHz. The distribution is strongly skewed: the median value is $0.86$~GHz, close to the crystalline reference of $0.73$~GHz, while individual values range from $0.18$~GHz (Sample~5) to $15.1$~GHz (Sample~6), spanning nearly two orders of magnitude. The substantially larger mean therefore arises from a few high-$E_J$ samples. This behavior is consistent with the exponential sensitivity of quantum tunneling to the local barrier thickness and atomic configuration: locally thinner or lower-barrier regions can produce much larger transmission and, consequently, much larger $E_J$. Because each amorphous atomistic model has a cross-sectional area of only $1.44\times1.25$~nm$^2$, such local structural fluctuations strongly affect the calculated values. In contrast, a realistic junction with a $200\times200$~nm$^2$ cross section samples more than $2\times10^4$ such microscopic regions, so these local variations are expected to be substantially averaged at the device scale.

\subsection{Area-normalized transmission}
\label{subsec:transmission}

\begin{figure}[!htb]
	\centering
	\includegraphics[width=0.62\linewidth]{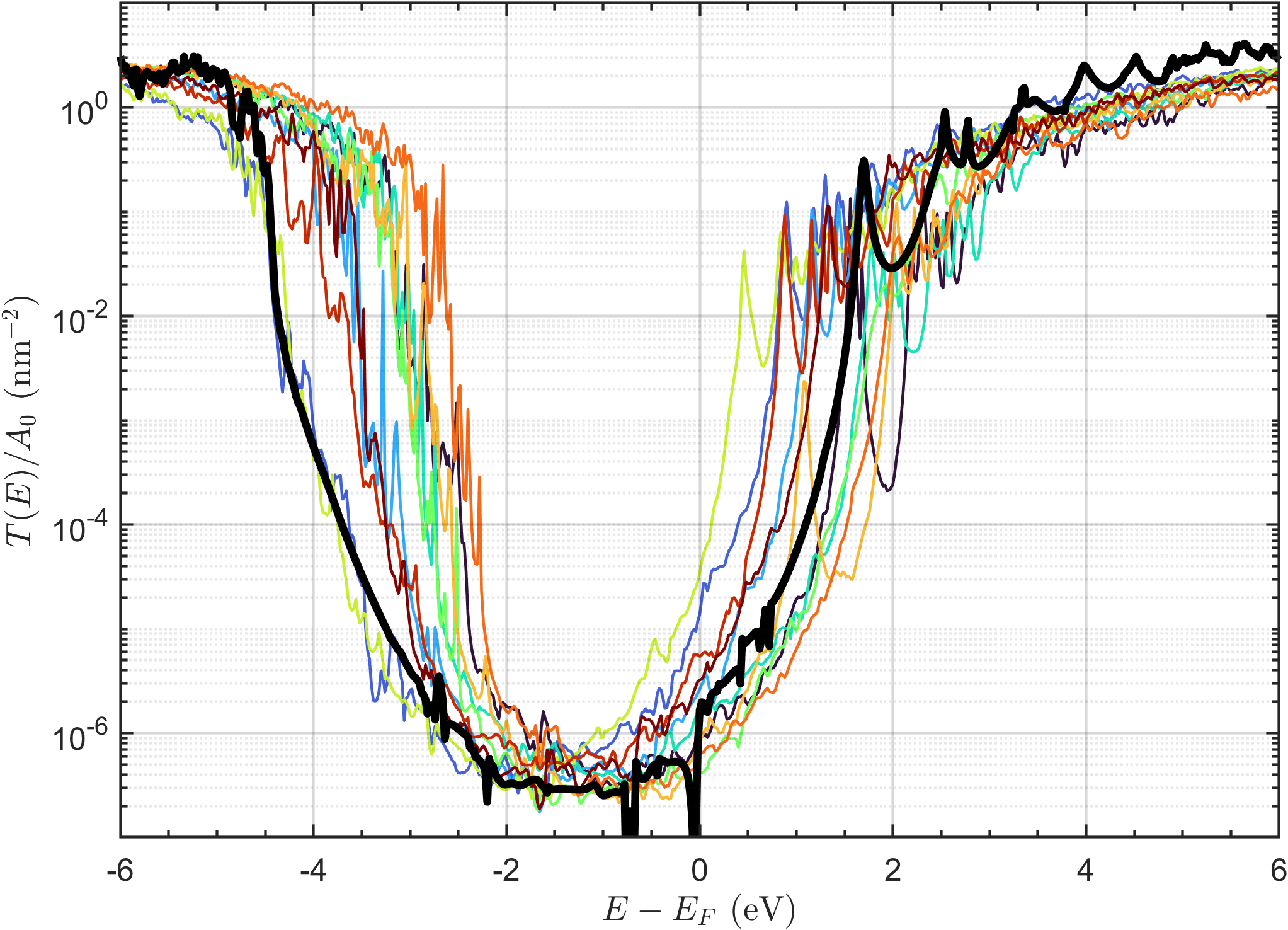}
	\caption{
		Area-normalized energy-dependent transmission $T(E)/A_0$ on a logarithmic scale, where $A_0$ is the transverse area of the corresponding simulation cell.
		The thick black curve is the crystalline junction; the colored curves are the ten amorphous junctions.
		Normalization by $A_0$ accounts for the different transverse supercell areas of the crystalline and amorphous models.
		All curves exhibit a similar transmission-valley structure, while their values near the Fermi level vary substantially among samples.
	}
	\label{fig:transmission}
\end{figure}

Figure~\ref{fig:transmission} compares the area-normalized energy-dependent transmission, $T(E)/A_0$, of the crystalline junction (black) and the ten amorphous junctions (colored), where $A_0$ is the transverse area of the corresponding simulation cell. This normalization accounts for the different transverse supercell areas required to preserve the periodic geometry of the crystalline and amorphous junctions. All curves share a similar overall shape: a deep transmission valley reflecting the insulating gap of the oxide, flanked by a steep rise as barrier states become available away from the gap. However, the position of the Fermi energy within this valley varies from sample to sample. Because the transmission varies by orders of magnitude across the valley, modest shifts of the valley relative to $E_F$ produce a large spread in the area-normalized Fermi-level transmission, $T(E_F)/A_0$, and consequently in the Josephson energies shown in Fig.~\ref{fig:ej}.

\subsection{Scattering states}
\label{subsec:scattering}

\begin{figure}[!htb]
	\centering
	\includegraphics[width=0.75\linewidth]{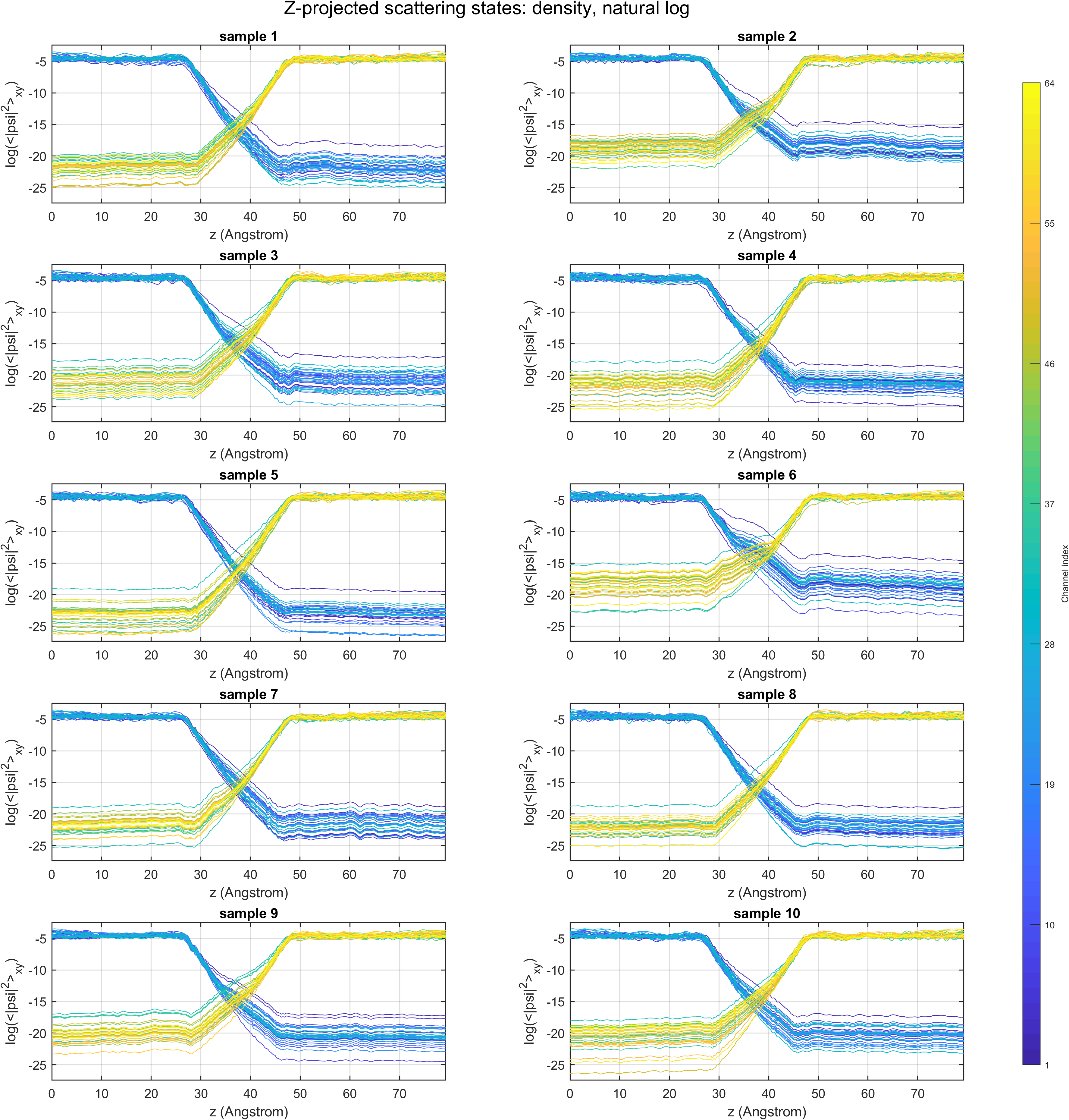}
	\caption{
		Probability of the Fermi-level scattering states (natural-log scale), averaged over the transverse $xy$ dimensions, for all 64 transport channels in each of the ten amorphous junction samples.
		The panels correspond to Samples~1--10, and the color denotes the transport-channel index.
		The approximately linear decay within the oxide on the logarithmic scale is characteristic of quantum tunneling; the different decay rates show that some channels are considerably more conductive than others.
	}
	\label{fig:scatt1d}
\end{figure}

To identify the microscopic transport mechanism, we analyze the Fermi-level scattering states. Figure~\ref{fig:scatt1d} shows, for each amorphous junction, the probability of the 64 scattering channels averaged over the transverse $xy$ dimensions on a logarithmic scale along the transport direction $z$. Inside the oxide the curves decay linearly on the log scale---the hallmark of evanescent decay---demonstrating that transport across the barrier is quantum tunneling. Channels incident from the left and right electrodes decay in opposite directions, producing the crossing pattern near the barrier. The decay rates differ among channels and samples: some channels are much more conductive than others, as evidenced by their slower decay within the oxide.

\begin{figure}[!htb]
	\centering
	\includegraphics[width=0.9\linewidth]{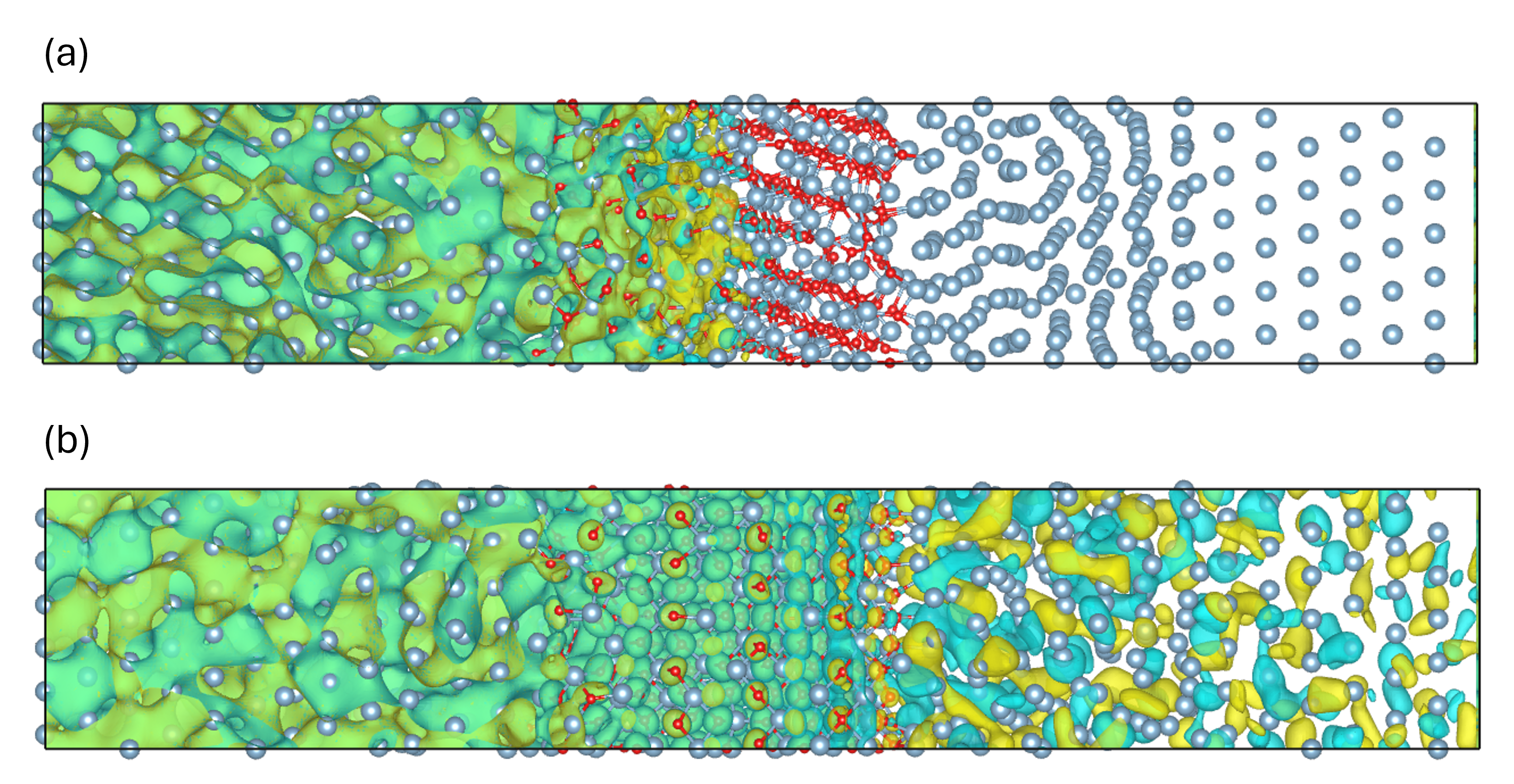}
	\caption{
		Real-space isosurfaces of two Fermi-level scattering states.
		(a) The least conductive channel of Sample~5.
		(b) The most conductive channel of Sample~6.
		Yellow and cyan isosurfaces denote positive and negative values of the wave function.
		The most conductive channel penetrates the oxide and remains extended on the far side, whereas the least conductive channel is localized near the incident electrode.
	}
	\label{fig:scatt3d}
\end{figure}

Figure~\ref{fig:scatt3d} visualizes two representative scattering states in real space: (a) the least conductive channel of Sample~5, the lowest-transmission junction, and (b) the most conductive channel of Sample~6, the highest-transmission junction. In the least conductive case, the wave function remains confined to the incident electrode and interface region and dies out inside the oxide; in the most conductive case, it penetrates the barrier and retains substantial amplitude on the transmitted side.

These observations suggest the following physical picture. Amorphous Al$_2$O$_3$ has a spatially nonuniform stoichiometry: Al-rich regions have a lower effective tunnel-barrier height, while O-rich regions have a higher one~\cite{fan2025,cyster2020}. When low-barrier regions happen to connect across the oxide, they form a percolation-like tunneling pathway, and electrons traverse the junction far more easily than through a uniform barrier of the same average composition; a similar formation of localized conduction channels driven by local density and stoichiometry variations was reported in atomistic NEGF simulations of AlO$_x$ tunnel junctions~\cite{cyster2020}. Samples with $E_J$ above the crystalline reference correspond to configurations with such a percolation path, and their dominant scattering states extend across the oxide, as in Fig.~\ref{fig:scatt3d}(b); samples with $E_J$ below the crystalline reference lack a connected low-barrier path, and their scattering states remain localized, as in Fig.~\ref{fig:scatt3d}(a).

\section{Conclusion}
\label{sec:conclusion}

We performed first-principles device modeling of Al/Al$_2$O$_3$/Al Josephson junctions with crystalline and amorphous tunnel barriers of the same oxide thickness. The ten amorphous junctions give a mean $E_J/h$ of $2.78$~GHz, with a sample standard deviation of $4.67$~GHz. The distribution is strongly skewed: its median of $0.86$~GHz is close to the crystalline reference of $0.73$~GHz, while a few high-$E_J$ samples increase the mean and the individual values span roughly two orders of magnitude. The large variance is explained by a percolation picture: stoichiometric inhomogeneity of the amorphous oxide creates Al-rich, low-barrier regions, and whenever these connect across the barrier the transmission is strongly enhanced, whereas samples without a percolation path can be even less transparent than the crystalline junction. A real junction with a $200\times200$~nm$^2$ cross section self-averages over more than $2\times10^4$ regions the size of our amorphous simulation cell ($1.44\times1.25$~nm$^2$), so its $E_J$ reflects the ensemble average of many microscopic configurations.

Two directions follow naturally from this work. First, computing a larger number of amorphous samples would enable a proper ensemble average and a statistical distribution of $E_J$, permitting direct comparison with measured junction-to-junction spreads. Second, the GGA functional used here underestimates the oxide band gap; employing GGA$+U$ or hybrid functionals such as HSE would yield a more accurate band gap and band alignment, and thus more quantitative values of the tunneling transmission.

\section*{Acknowledgments}
We thank the Digital Research Alliance of Canada for the computational facilities that made this work possible. 
H.G. acknowledges support from the Natural Sciences and Engineering Research Council of Canada (NSERC).
We are grateful to Prof.\ Nicholas C. Strandwitz for sharing the amorphous atomic structures used as references.


\end{document}